\documentclass[12pt]{article}
\usepackage{graphicx}
\begin{document}
\centerline{Boundary effects in a three-state modified voter model for
languages}

\bigskip

\centerline{T. Hadzibeganovic$^1$, D. Stauffer$^2$, C. Schulze$^2$}

\bigskip
\noindent
$^1$ Language Development \& Cognitive Science Unit,
Karl-Franzens-University Graz, Heinrichstr. 36, 2. OG; A-8010 Graz, Euroland

\bigskip
\noindent
$^2$ Institute for Theoretical Physics, Cologne University, \\
D-50923 K\"oln, Euroland

\bigskip
\noindent
e-mail: stauffer@thp.uni-koeln.de, ta.hadzibeganovic@uni-graz.at

\bigskip
The standard three-state voter model is enlarged by including the outside 
pressure favouring one of the three choices and by adding some biased internal random noise. 
The Monte Carlo simulations are motivated by states with the population divided into three 
groups of various affinities to each other. We show the crucial influence of 
the boundaries for moderate lattice sizes like $500 \times 500$. By removing 
the fixed boundary at one side, we demonstrate that this can lead to the victory 
of one single choice. Noise in contrast stabilizes the choices of all three populations. 
In addition, we compute the persistence probability, i.e., the number of sites who have 
never changed their opinion during the simulation, and we consider the case 
of "rigid-minded" decision makers. 
   
\bigskip
\section{Motivation and Model}
The political situation after the break-up of former communist powers and the 
emergence of new sovereign states in Europe and elsewhere, justify yet another 
look at linguistic practices as informed by geopolitical agendas with the ever 
growing asymmetric power relations and the ongoing struggle for the accumulation 
of linguistic and cultural capital. Pronounced language asymmetries with highly 
competitive behaviour have caused a situation in which no successor state can claim 
a one-and-only homogenous "national language" without serious caveats. In complex 
decision making, there are typically no single agreements when large numbers of 
decision makers are expected to choose from a large set of alternatives \cite{jones}. 
Any attempts to tackle the intricacies of these phenomena at the more global level, 
stumble across a series of theoretical and methodological problems. However, more 
recent studies have shown that if individuals tend to share similar knowledge structures 
within a given choice domain, then a rather stable global choice behaviour is observed 
with about 90\% probability \cite{richards}. 

In the present study, we shift the analysis of geographical \cite{lim}, linguistic, 
sociological and political factors to another report \cite{hadz}, while looking here 
for a model which may describe a type of language competition observed in an environment 
populated by strong minorities facing several alternative choices
 and partially bordering 
on supporting states. We do not intend to discuss whether "dialects" would be a better 
name instead of "languages", however, we notice that recent linguistic 
analyses \cite{moenneslanda} could not trace any dialectal differences following 
national lines in many of the successor states in e.g. Southeast Europe. 
Instead, it has been increasingly argued \cite{moenneslandb} that all different 
groups in the region tend to use exactly the same idiom.  However, the restructured 
political pictures lead to the emergence of completely new policies, such that the 
question of language has become a top political issue in a community which is 
linguistically homogenous but politically divided \cite{moenneslandb}. We treat this 
problem as the one of opinion dynamics where everybody can adopt one of the three choices 
A, B, and C (each representing the opinion about the linguistic identity), with transitive  \cite{houston}
preference relations. Thus we model the evolution of the global choice behaviour in a 
tripartite system where due to particular economic and political alliances, 
languages may happen to be in a closer contact at one point in time and more divided 
at another. As a consequence, people may start adopting linguistic features or even 
full languages of their neighbours, if they have sufficient gains or are influenced 
by a set of social and/or political factors. This is especially valid for those 
languages which both belong to the same linguistic family and border with one another. 
 
In our model, we assume that the simulated $L \times L$ lattice is bordered on top by 
the population preferring the linguistic choice A and on bottom by the population preferring 
the choice C. This we achieve through two boundary lines of only A on top and only C on 
bottom. Initially, in the middle of the lattice, the decision makers (DM) preferring 
the choice B are dominant, while on the top we have DM mostly preferring A and on the 
bottom mostly preferring C, with the concentrations of A, B, and C varying linearly 
with height. Figure 1 shows for $L = 1000$ the initial distribution of choices 
A, B, and C as a function of height, i.e., we plot the numbers of A choices in each 
horizontal line while doing the same for the B and C choices.

\begin{figure}
\begin{center}
\includegraphics[angle=-90,scale=0.4]{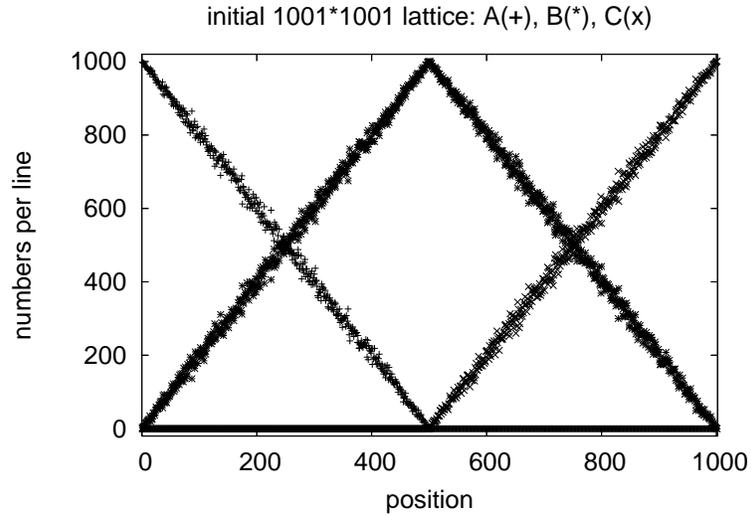}
\end{center}
\caption{
Initial distribution of A, B and C  as a function of height, with
concentrations varying linearly with height. Half of the people selects the B choice (mostly
in the middle), one quarter selects A (mostly near one border) and one quarter 
selects the choice C (mostly near the other border).
}
\end{figure}

\begin{figure}
\begin{center}
\includegraphics[angle=-90,scale=0.4]{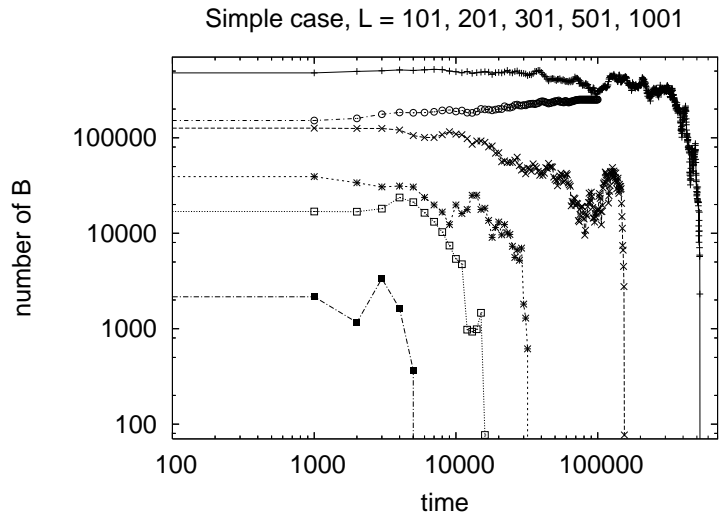}
\end{center}
\caption{
Double-logarithmic plot of B choices versus time in the simple voter model.
}
\end{figure}

\begin{figure}
\begin{center}
\includegraphics[angle=-90,scale=0.4]{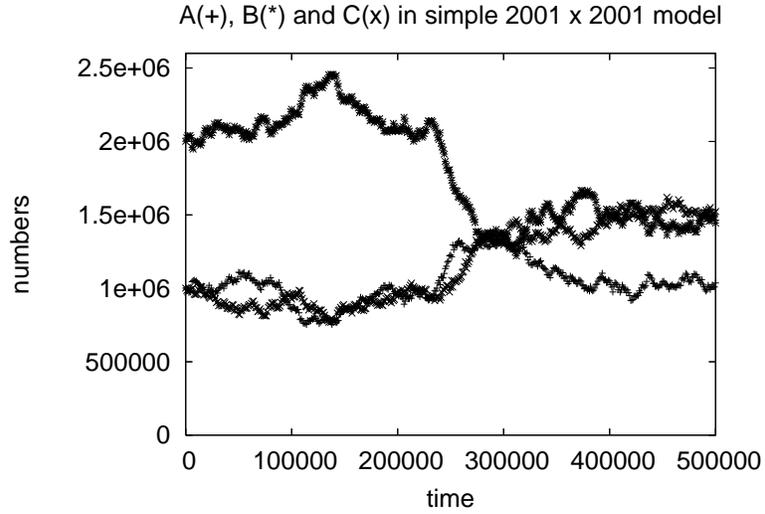}
\end{center}
\caption{
All three choices for the largest lattice in the simple voter model.
}
\end{figure}

\begin{figure}
\begin{center}
\includegraphics[angle=-90,scale=0.4]{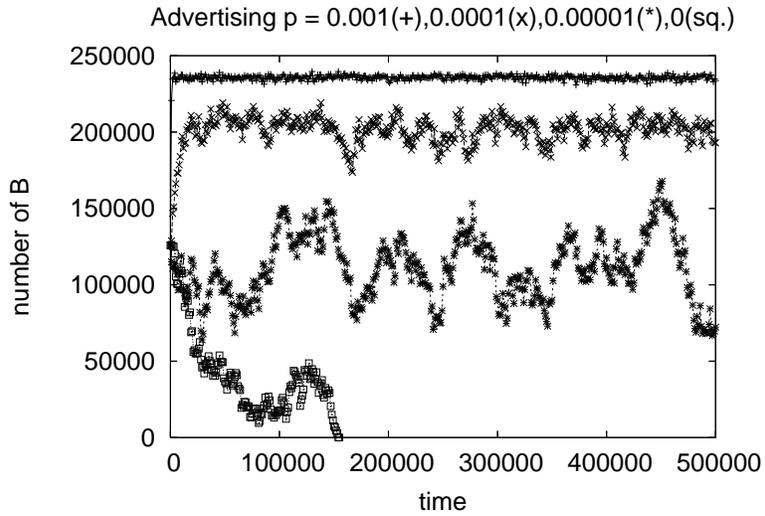}
\end{center}
\caption{
A small amount of "advertising" $p$ in favour of B stabilises the B choice;
a larger $p$  increases it, though not to 100 percent = 251001.  No "noise" is 
included yet.
}
\end{figure}

\begin{figure}
\begin{center}
\includegraphics[angle=-90,scale=0.4]{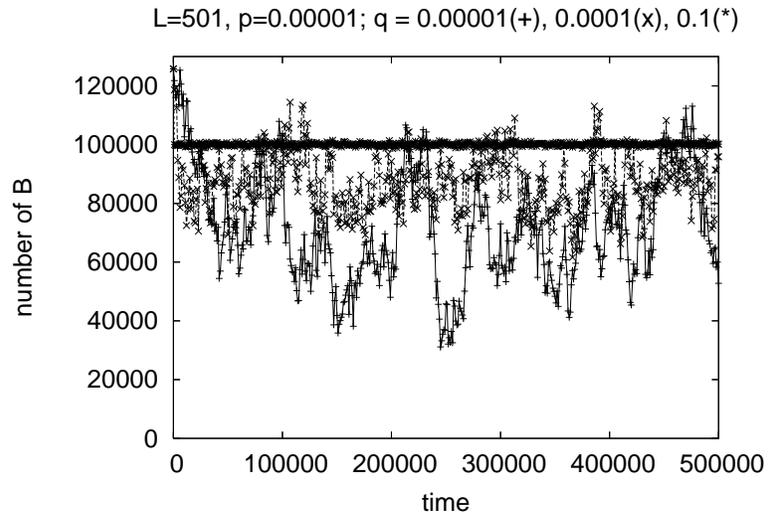}
\end{center}
\caption{
Addition of noise $q$ to voter model plus advertising reduces fluctuations.
}
\end{figure}

\begin{figure}
\begin{center}
\includegraphics[angle=-90,scale=0.4]{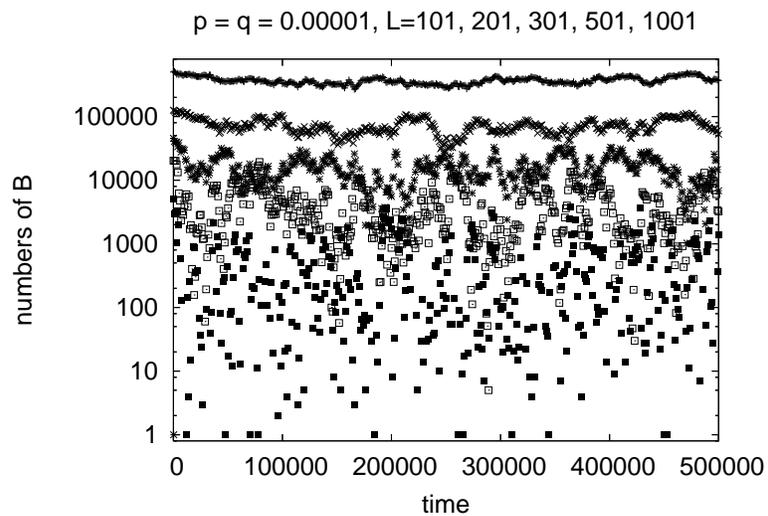}
\end{center}
\caption{
Dependence of B choice on lattice size, with small advertising and small 
noise.
}
\end{figure}

\begin{figure}
\begin{center}
\includegraphics[angle=-90,scale=0.4]{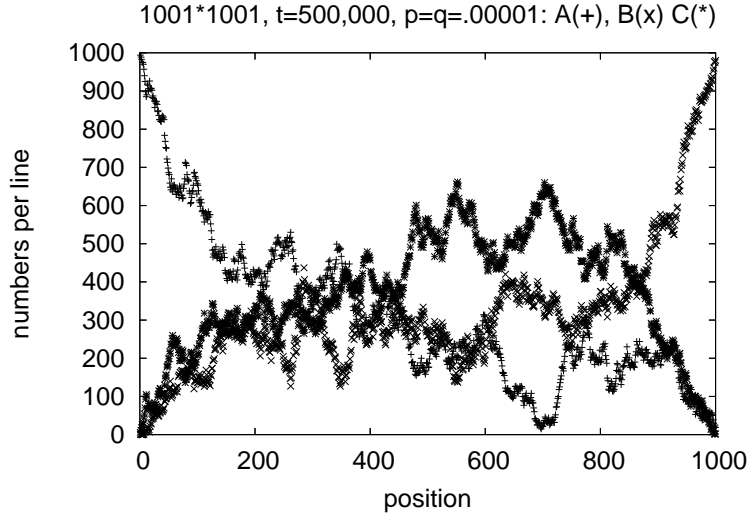}
\end{center}
\caption{
Geographic distribution of A, B and C opinions after equilibration as a 
function of the line number in the lattice; $p = q = 10^{-5}$. 
}
\end{figure}

\begin{figure}
\begin{center}
\includegraphics[angle=-90,scale=0.26]{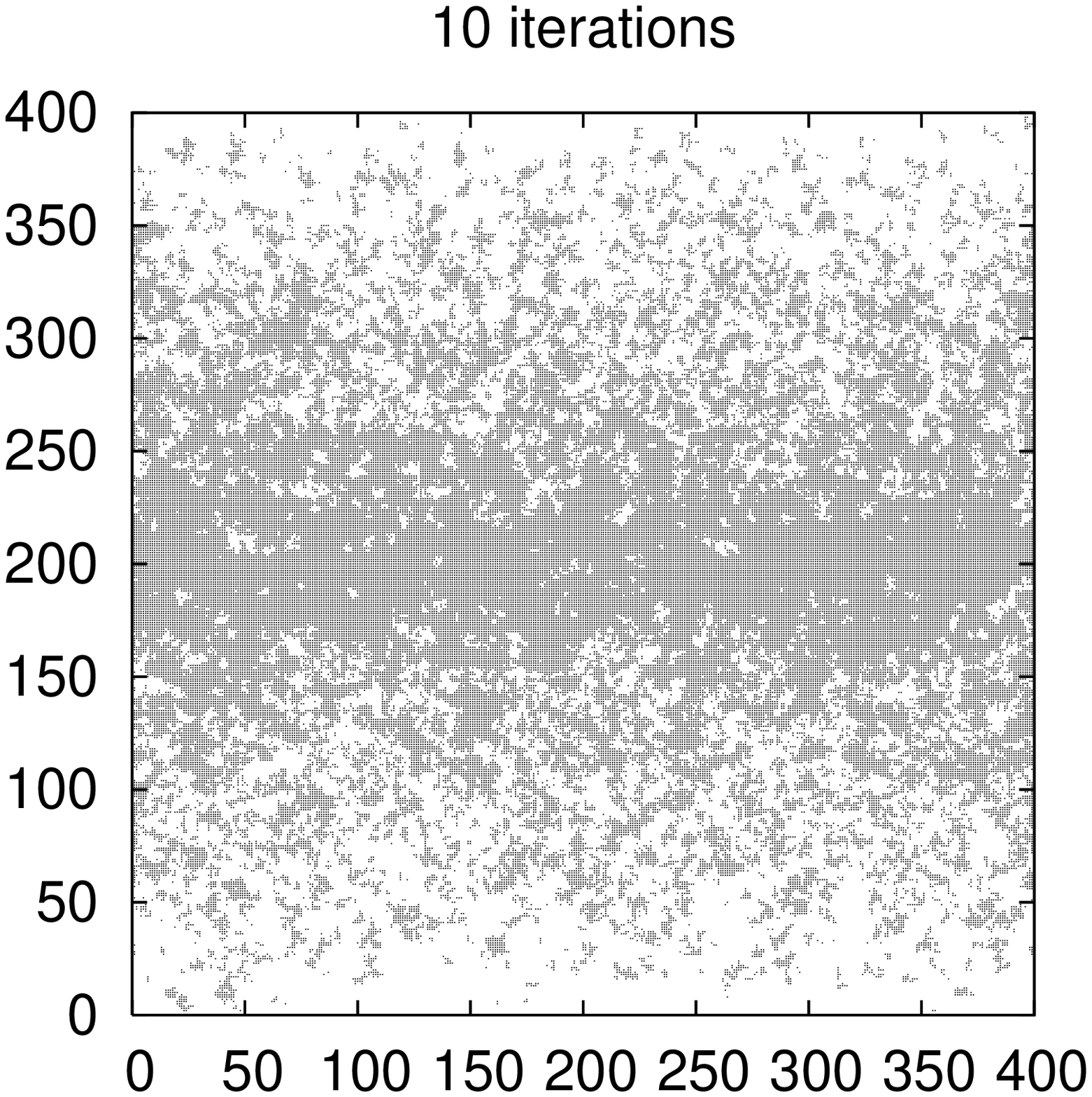}
\includegraphics[angle=-90,scale=0.26]{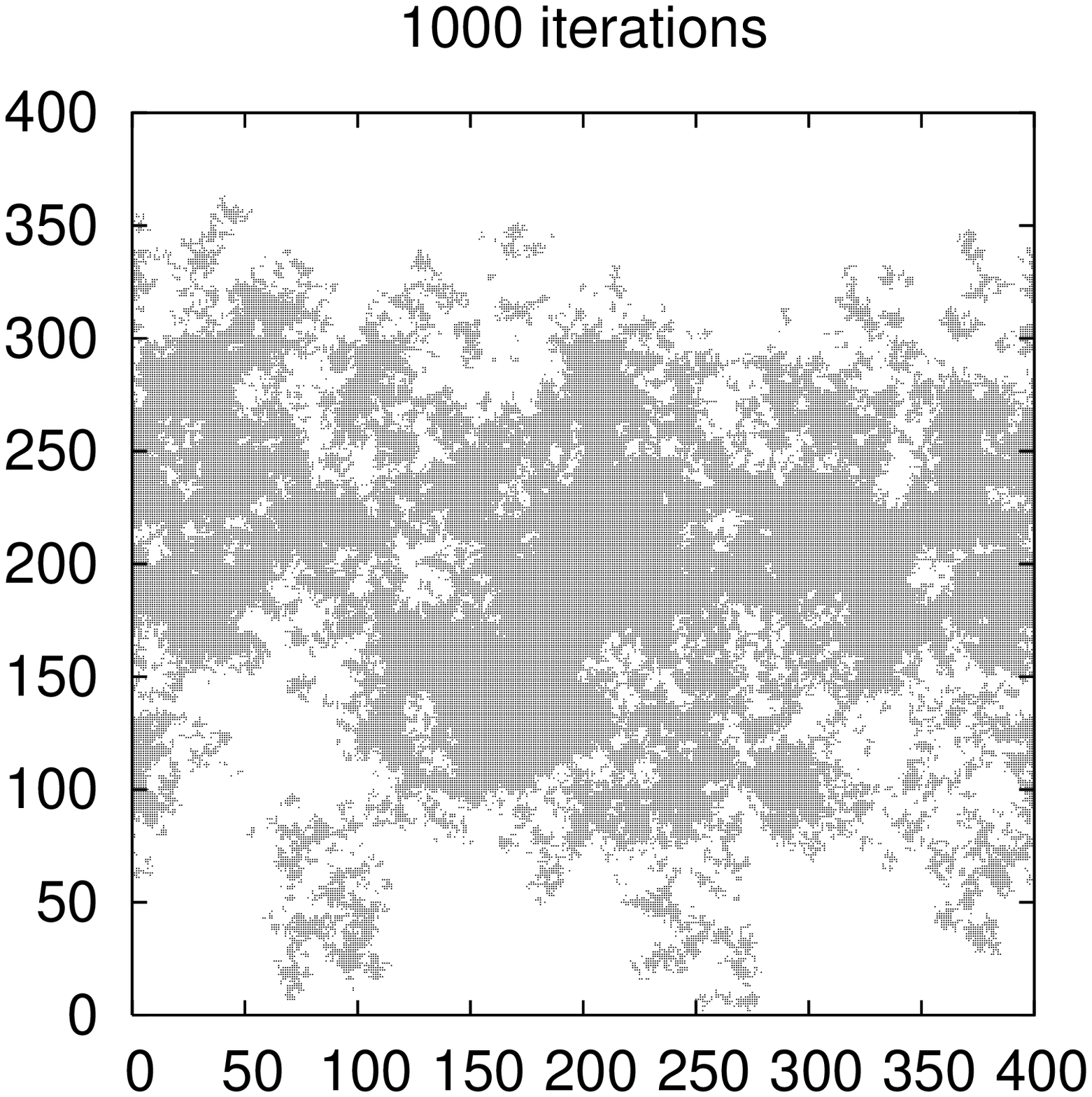}
\includegraphics[angle=-90,scale=0.26]{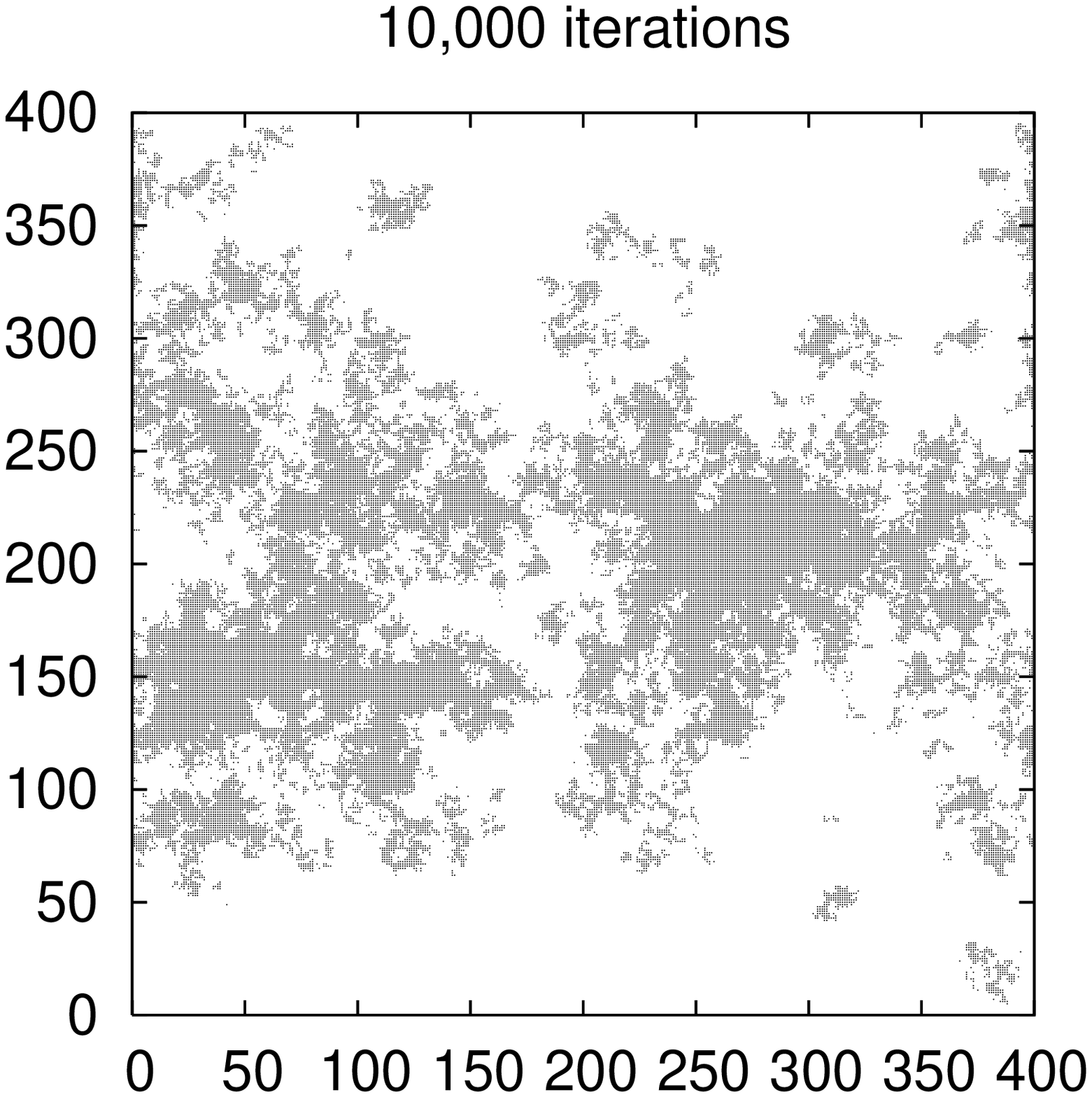}
\includegraphics[angle=-90,scale=0.26]{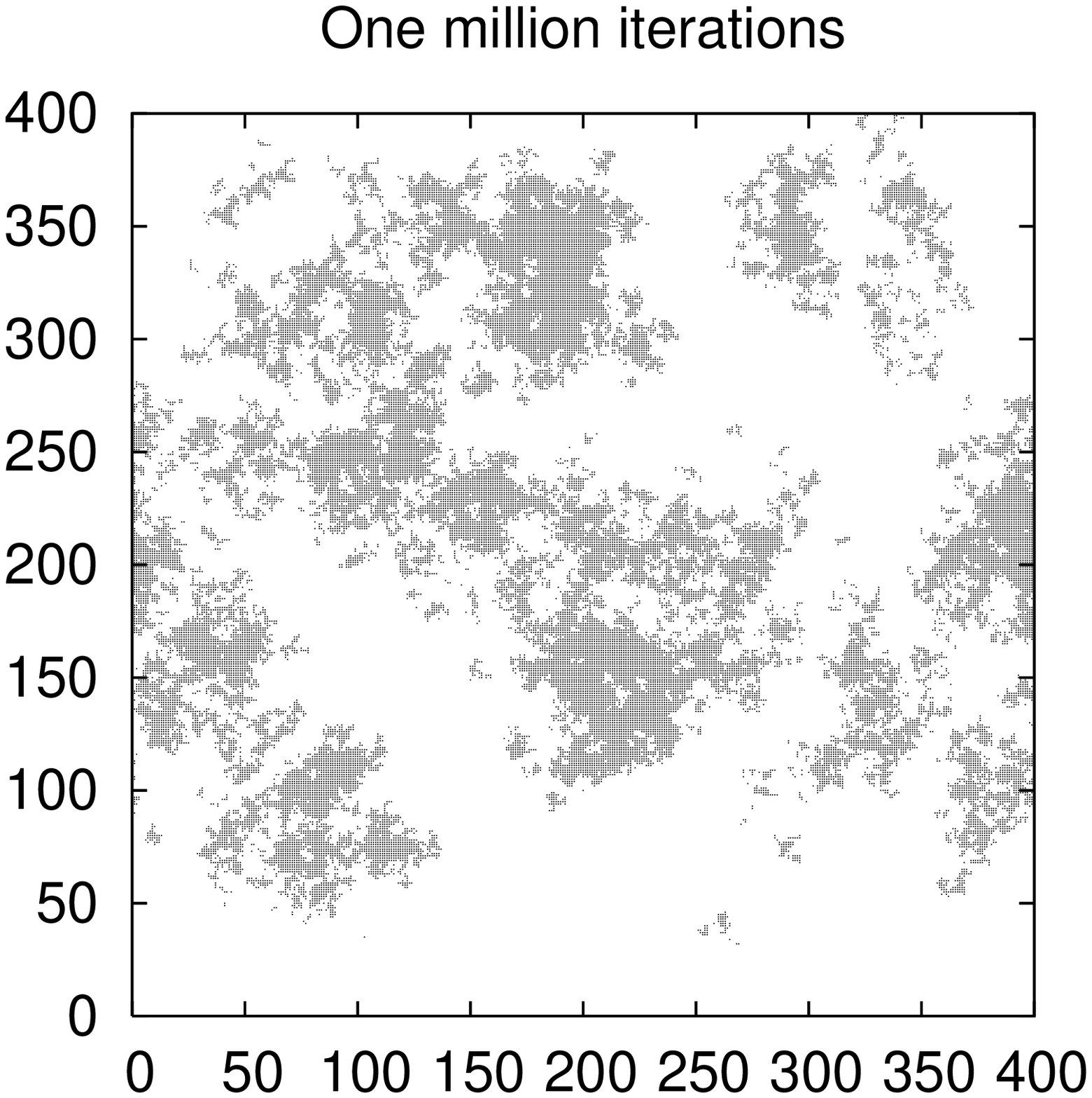}
\end{center}
\caption{
Two-dimensional distribution of the B opinion as a function of time, with 
small noise and small advertising, $p = q = 10^{-5}$.
}
\end{figure}

\begin{figure}
\begin{center}
\includegraphics[angle=-90,scale=0.4]{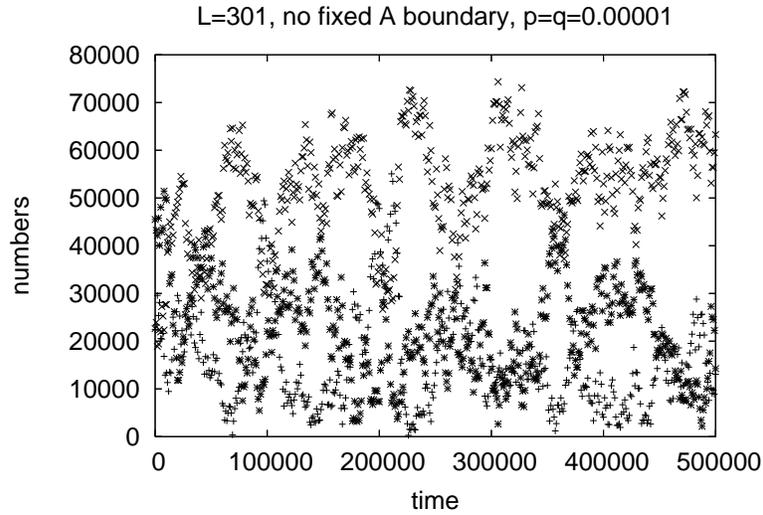}
\end{center}
\caption{
C is likely to win if rigid A boundary at A side is removed. For smaller $L$ this
victory is more pronounced, for larger $L$ less.
}
\end{figure}

\begin{figure}
\begin{center}
\includegraphics[angle=-90,scale=0.4]{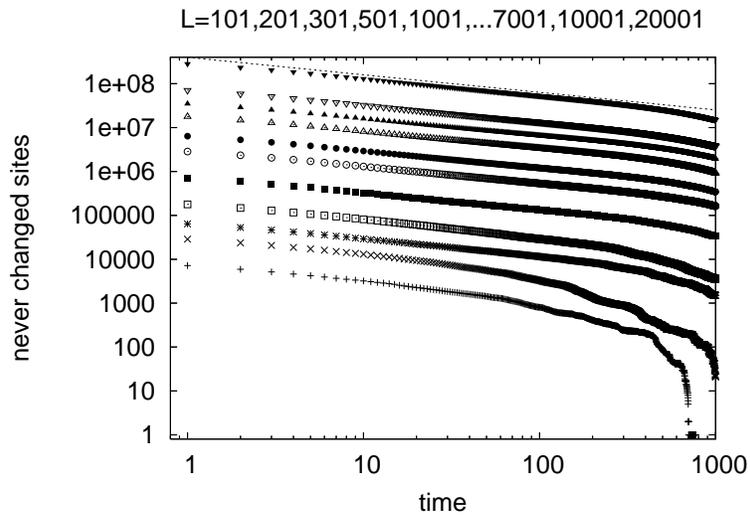}
\end{center}
\caption{
Persistence: Number of sites which have not changed since the beginning of the 
simulation. For longer times this number decays faster, reaching zero before
or near 20,000 iterations.  
}
\end{figure}

\begin{figure}
\begin{center}
\includegraphics[angle=-90,scale=0.4]{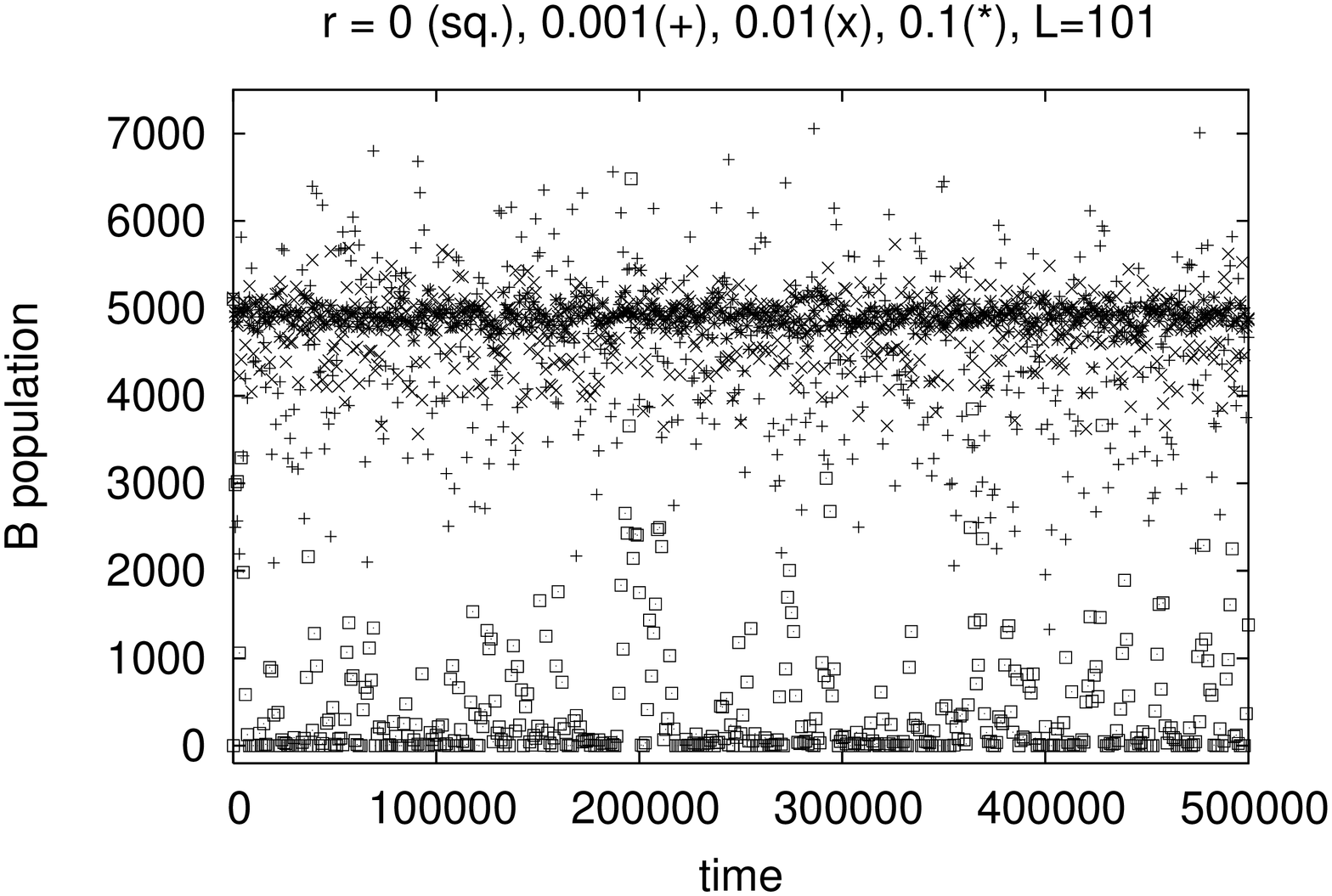}
\end{center}
\caption{
Persistence: B choice if a random quenched fraction of 0.1, 1 and 10 \% 
never changes in a $101 \times 101$ lattice; $p=q=10^{-5}$. For comparison
we repeat from Fig. 6 the data without such a fraction (squares at the bottom). 
}
\end{figure}

Opinion dynamics can be described in several different ways, by using e.g. 
the models of Deffuant et al. \cite{deff}, Krause and Hegselmann \cite{krause}, Sznajd \cite{sznajd}, 
or some older approaches such as the model of Axelrod \cite{axelrod}. In the present paper, 
the basic dynamics is the traditional voter model \cite{sanmiguel}, where at each iteration every site 
accepts the language of a randomly selected nearest neighbour. Then every individual 
with choice A having three individuals with choice B and one with A as neighbours is 
likely to give up the choice A and shift to B. Instead of this voter model we also could 
have used a Potts-type model \cite{schulze}, as did Lim et al. \cite{lim} without mentioning Potts, or an 
Ising model with spin $+1, 0, -1$. The voter model, however, seems to be simpler, and is a 
first step towards a more complete future work on this topic.

In addition, we assume that with low probability $p$ at each iteration each 
individual accepts the choice B because of "advertising campaigns" \cite{schulzead} from 
higher authority. Furthermore, we added noise to the whole process, in order to 
mimic perturbations and random fluctuations in a person's switching behaviour 
among choice items over time due to e.g. mixed marriages, sudden preference changes of latent origin, 
or other personal reasons of individual DM. 
More precisely, with low probability $q$ each individual at each iteration shifts to 
another randomly selected choice: C shifts equally often to A and B, B shifts to A 
three times more often than to C, and A shifts to B three times more often than to C. 
We use this particular update rule for the opinion dynamics because we wish to model a 
situation in which there is a virtual i.e., political barrier splitting the lattice into 
two major pieces such that C dominates on one side of the barrier, while A and B are 
predominantly preferred on the other (e.g. through the separation into two political entities). 
Furthermore, each of the three choices A, B, and C has some special property: 
Only the population with B choice has no border with fixed sites, while populations preferring 
A and C choices partially border on two areas from which they sporadically obtain additional 
political support. In addition, C choice is rather disliked by those individuals choosing A and B. 
Thus, we assume that individuals' information is not arbitrary but is constrained by the 
individuals' context-based mental model of the respective choice domain. 

We further investigate the case in which one of the two fixed boundaries in the model 
is removed (e.g. when the neighbouring side weakens or disrupts its support), 
assuming that this might lead to the victory of one of the opinion alternatives within the tripartite system.  
We also compute the persistence probability $s(k, t)$ for different lattice sizes, where $k$ is the number 
of opinions in the original voter model, and $p = q = 0.00001$.

We do not take into account voluntary or forced migration of DM; neither 
have we included the factor proportional to the population size dynamics (birth vs. death rates). 
Of course, the question of how exactly individuals aggregate their preferences over $k$ candidate 
languages involves much more fine-granularity than considered in the present model. 
Indeed, the opinion switching dynamics might be far more complex than described 
by the update rule in the present study. For instance, not only single individuals, but moreover, 
larger amounts of population members might drastically change their attitudes towards 
linguistic (and political) campaigns such that they no longer participate in the opinion 
switching process. Thus, they might just "tune out" \cite{fonghsu} at one or more points in time, 
thereby considerably affecting the overall evolution of preferences or choices. 
At such times, the individual state remains frozen, until a particular event 
causes the decision maker to re-enter the switching process. In order to reflect on this type of 
behaviour in our simulations, we included a situation in which a particular random (quenched) fraction 
of population never changes its opinion (fixed set of "stubborn" decision makers). However, we assume here 
that some aspects of the human opinion dynamics are independent of the detailed micro-processes 
of decision making, relying more on some overall, universal properties. These microscopic 
quantities are besides not always directly accessible to the observer. On the other hand, 
there are macroscopic quantities, defined by parameters that are fixed from the outside. 
The main goal of statistical physics is exactly to provide a link between these microscopic 
and the more global, macroscopic aspects of an investigated system.

\section{Results}

For the simple voter model without advertising and noise, Fig.2 shows how the choice B 
decays due to the influence of the fixed boundaries with A and C only. The larger the lattice, 
the longer does the B choice survive. The single increasing curve for $L = 501$ shows the B choice 
if the two boundary lines are all B, instead of all A and all C, respectively. 
This figure thus indicates the importance of the lattice boundaries. For the same simple model 
but a larger lattice, Fig.3 shows the time dependence of the A, B and C opinions. 

If advertising in favour of B is added to the voter model, the B opinion can be prevented 
from dying out and can even increase, enlarging thus its initial majority; see Fig.4. 
Very small advertising probabilities $p > 10^5$ suffice. 

All following figures include both advertising with $p = 10^{-5}$ and noise 
with varying small $q$. Figure 5 shows that somewhat counter-intuitively, 
the noise reduces the fluctuations if its strength $q$ increases. For $q = 0.1$ already the 
results agree well with theoretical expectations of 40 \% A, 40\% B, 20\% C, 
the stationary solution of the rate equations for noise only. Again, small lattice sizes let 
the B opinion decay; however now it happens that after B has died out it is resurrected 
via advertising and noise; see Fig.6. Fig.7 shows what happens in the long run with the initial 
distribution of A, B and C, as a function of the line number of the lattice. 
Fig.8 shows as a function of time how the voter process first leads to clustering: 
A prefers to be near A, B near B, and C near C. Later, the decrease in preference for 
the B opinion is seen until it is stabilized by noise and advertising. 

If the rigid boundary of 
pure A at one end is omitted, then both B and A are strongly reduced and C wins for small $L$; see Fig.9. 

The number of sites which never changed during a simulation decays initially
as about $1/t^{0.4}$ for $t < 10^3$, and then stronger down to zero, see Fig.
10 and straight line there. This power law differs from the behaviour of
other voter models with different noise or without noise \cite{voter}.

If we assume a small fraction $r$ of sites never to change, then they 
help the B population to survive better, and for larger $r$ they reduce
fluctuations, as seen in Fig.11.

\section{Discussion} 

In the present paper, we proposed the generalization of the standard three-state voter model 
for a linguistically homogenous but politically divided population with subgroups of different 
mutual affinities. Starting with this tripartite initial configuration, the Monte Carlo simulations 
were conducted for different lattice sizes, different amounts of "advertising" (outside pressure) 
and asymmetric noise distribution. These simulations lead us to several, non-trivial conclusions. 

First, the individual preferences for the initially most dominant opinion are more prone to decay 
given the rather unstable and fuzzy borders with respect to the two other populations with more 
fixed boundaries and opinions different from the majority. Thus, we found a substantial importance 
of the boundary effects for the investigated population when placed on smaller lattices. However, 
increasing the size of the lattice prolongs the survival of the most dominant opinion. 

Second, adding some "advertising" in favour of the dominance state prevents the opinion of the 
majority from extinction, while considerably increasing its size. In the model, the "advertising" 
probabilities refer to the different outside pressures of e.g. the international community in favor of the largest 
population due to particular reasons. Somewhat counter-intuitively, when noise was added to the process, 
it reduced the previously observed fluctuations in choice behaviour. The noise parameter in our model 
can relate to a set of different factors such as marriages and relationships between culturally and 
linguistically different individuals, unexpected changes of preferences for different political parties 
with different language policies, or other personal reasons of individual decision makers. 

In addition, we investigated a particular case of removing a rigid boundary at one of the previously stable sides, 
thus yielding the victory of only one single opinion for small lattice sizes ($L = 301$). The decay function for 
individuals who never changed their opinion during the simulations, tends to follow a power law which differs 
considerably from the behaviour documented in other studies of persistence probabilities, where different types of noise 
were used or no noise was applied at all [15]. Finally, if for some reasons a particular fraction $r$ 
of population "tunes out" from any further opinion switching, this leads then to a better than the previously 
observed stabilization of the most dominant state, with additional reductions of fluctuations for larger $r$. 

The update rule reported in this study was used in order to mimic situations in which three populations are 
not just geographically, but also virtually, i.e., politically divided (e.g., via formation of two separate 
entities following an inter-cultural conflict). Further research in this direction should therefore include a 
situation where the political barrier is removed, thus necessitating different update rules in the model. 

Our investigation was motivated by a phenomenon in which three groups of a given population speak one and 
the same language (linguistic homogeneity), but have different opinions about their own linguistic identity 
(political diversity), thus each claiming to speak different language from the "other". In such situations, 
the name which people usually attach to "their language" is the one of their supporting neighbouring or other states. 
As a consequence, individuals can increasingly start to adopt linguistic features from languages of their 
supporting neighbours, especially if they were previously in a conflict with their within-population neighbouring 
members. If no supporting neighbouring states are available,
%(i.e., the subpopulation is fully occupied by two other or even more subpopulations), 
people will try to invent or dig up half-forgotten words that function as shibboleths 
(dialect identifiers). Eventually, certain aspects of pronunciation that people are conscious of could be 
changed "by decision", while grammar is expected to remain unaffected, since people are not conscious of most 
of the grammar they use. Indeed, as recent scientific studies have shown, an immensely large fraction of our 
everyday behaviour is unconscious, i.e., zombie-like \cite{koch}, and consequentially, under rather uncertain 
conditions, we tend to act more as irrational decision makers \cite{kahnemann}. For instance, one opinion might 
often be preferred to another, even if the probability of yielding the worse outcome with it is one. 
Often irrational and particularly politically-driven "swings" cause people to make more (unnecessary) 
decisions about their language and the language of others. From the linguistic perspective, one can predict 
in the aforementioned case that only highly salient linguistic features can change more rapidly. 

People can start adopting other linguistic features or even fully switch to other languages due to many reasons: 
due to the use of different languages for different purposes (diglossia), because they decide to migrate to 
a foreign country, because the status of the non-native language is higher, or if their country is conquered 
by the linguistically different population. However, it was demonstrated that contrary to predictions of some 
models \cite{abrams}, a stable co-existence of languages in competition, regardless of their status, is still 
possible if they are sufficiently similar to each other \cite{mira}. Typical instances of such "linguistic areas" 
where several languages compete and share a number of features are found in Southeast Europe [4]. For example, 
in the region of Sanjak, several different languages share many of their features. The area of Sanjak is 
politically divided between the states of Serbia and Montenegro, and geographically and linguistically surrounded 
by four different states or areas (Serbia, Montenegro, Bosnia-Herzegovina, Kosovo-Metohia). In parts with Bosniak 
population in Sanjak, individual speakers claim to speak Bosnian language, however, the laws in Serbia allow Bosnian 
only as an "elective" in primary and secondary education, while Montenegrins "solved" this "problem" by 
labeling the language exclusively as "mother tongue" without any reference to any particular national language. 
Most recently, the new Montenegrin constitution introduced the new and the official Montenegrin language in the country, 
with other languages (Bosnian, Serbian, Albanian and Croatian) as additional, but secondary official languages 
of the state. This was preceded by a high disagreement about how the official language should be called, with competing 
opinions between those who viewed themselves as of different cultural and ethnic origins. 

In Bosnia-Herzegovina, we have a related case, where strong minorities decided to introduce the standard languages 
of the neighbouring countries as their official idioms [6]. It seems therefore that individuals, and more importantly, 
much larger population clusters, indeed tend to consciously introduce several novel features in their languages 
just to distinguish themselves from "others". As we have found in our simulations, the final outcome of this process 
might strongly depend upon the character of the boundaries defined between different areas, but also on the size of 
the populated region, the internal random noise, and the outside pressure from some higher authority. We also wish 
to stress at this point that despite of the obvious structural similarity and mutual intelligibility of many of the 
Southeast European languages, these two factors should not be taken as decisive (or misused) in language planning 
or policy making, especially because some other world languages are mutually intelligible to a large degree just as 
e.g. Croatian and Bosnian (such as Urdu and Hindi), but are recognized as separate standard languages.

A counterexample to the Bosnian case would be Estonian, where the placement of verbs at the end of sentences 
was {\it consciously avoided} in order to differentiate the language from German, a language which it had been 
under the influence of. But this was of course an attempt guided by linguists. It is in fact not quite clear whether 
the word order had been influenced from German or whether verb-final order was just a natural, internal Estonian 
development. Somewhat more cooperatively, in the northwest Amazon region, there is a marriage system according to 
which people have to marry someone speaking a language that is different from their own: linguistic exogamy. Thus, the 
language which defines your identity is that of your father. In this situation people are careful not to borrow 
words from other languages, but they have influenced one another at the level of grammar to a great extent. 

Explanations for provocative phenomena of this kind are best investigated within the framework of opinion dynamics; 
the voter model is a traditional example and is a good place to begin. We believe that further refinements and 
applications of our generalized voter model, carried over more realistic input while simultaneously considering 
a more complex set of intervening variables, might tease apart the multiple forces (latent) and practices (observables) 
through which individual decision makers shift to other languages or opinions. This can further shed light on the 
related psychological mechanisms responsible for integration and preference dynamics of these practices. For example, 
from the psychological point of view, one can investigate whether some collective phenomena of this type are better 
explained by means of modeling the individual {\it indifference classes} rather than {\it preferences} among 
various choices \cite{alcantud}. Thus, in rather conflicting and culturally highly mixed areas, agents are 
usually better aware of what they don't like, then what would they prefer more. 

Additionally, more degrees of freedom of individual "voters" (using complex networks instead of lattices, 
allowing for a co-evolution of the network and voter states), as well as migratory processes and population 
size dynamics should be incorporated in future modeling. It would also be ideal to integrate the opinion 
dynamics models with those of language evolution and learning \cite{niyogi,ssw}. Models of opinion dynamics can 
also help in future conflict resolution in the aforementioned regions, by suggesting e.g. possible modes of "softer" 
and scientifically-valid language policies, or by (re)defining particular linguistic norms as opposed to those driven 
by political factors. 

We finally argue that the advantages of our generalized three-state voter model span beyond mere linguistic applications, 
since the model is able to simulate and characterize the nature of both collective political and linguistic 
states (i.e. their co-evolution). Moreover, through future comparison with realistic data, we might be able {\it to predict} 
whether and under which conditions various highly-similar languages can stably co-exist, and whether and how the 
political stability in a given region changes with the ever growing "linguistic diversity". Physicists and other 
computational modelers can significantly contribute to this research domain, by developing tools capable of establishing 
more isomorphic relations between the model parameters and the realistic traits of languages or opinions in competition.

\section{Acknowledgements}

We thank  S. Wichmann for related information on Estonian, the linguistic practices of the 
northwest Amazon region and the role of conscious agent behaviour in language change processes.

\end{document}